\documentclass[reprint, secnumarabic, amssymb, amsmath, superscriptaddress, aps, prl]{revtex4-2}
\usepackage{graphicx}
\usepackage[colorlinks=true,urlcolor=blue]{hyperref}
\usepackage{xcolor}

\hypersetup{
colorlinks,
linkcolor={red!50!black},
citecolor={blue!50!black},
urlcolor={blue!80!black}
}
\usepackage[normalem]{ulem}
\usepackage{units}

\begin{document}

\title{Quantum entanglement and quantum geometry measured with inelastic X-ray scattering}

\author{David Bałut}
\affiliation{Department of Physics, University of Illinois, Urbana, Illinois 61801}
\affiliation{Materials Research Laboratory, University of Illinois, Urbana, Illinois 61801}

\author{Barry Bradlyn}
\affiliation{Department of Physics, University of Illinois, Urbana, Illinois 61801}
\affiliation{Anthony J Leggett Institute for Condensed Matter Theory, University of Illinois, Urbana, Illinois 61801}

\author{Peter Abbamonte}
\affiliation{Department of Physics, University of Illinois, Urbana, Illinois 61801}
\affiliation{Materials Research Laboratory, University of Illinois, Urbana, Illinois 61801}
\email{abbamonte@mrl.illinois.edu}

\begin{abstract}

Using inelastic X-ray scattering (IXS), we experimentally investigate the quantum geometry and quantum information in the large-gap insulator, LiF. Using sum rules for the density-density response function measured in IXS, we compute the quantum Fisher information of the equilibrium density matrix of LiF associated with density perturbations. Next, by exploiting universal relations between the quantum Fisher information, the optical conductivity, and the quantum metric tensor, we extrapolate the diagonal $(h,k,l) = (1,0,0)$ component of the quantum metric of LiF, known as the quantum weight. We compare our results to recently-proposed bounds on the quantum weight and find that the quantum weight in LiF comes close to saturating a theoretical upper bound, showing that quantum-mechanical delocalization plays an important role even in ionic insulators. Our work serves as a proof-of-principle that IXS techniques can be used to quantify state-of-the-art quantum geometric quantities of materials, and establishes the quantum Fisher information as an experimentally-accessible generalization of quantum geometry to real materials.

\end{abstract}

\maketitle

%intro
\section{Introduction}

Recent advances in quantum materials have sparked renewed interest in the study of the geometric properties of quantum states~\cite{provost1980riemannian,torma2022superconductivity,torma2023essay,jackson2015geometric,roy2014band,verma2024instantaneous,komissarov2024quantum,wang2021exact,fang2024quantum,onishifundamentalbound,onishiUniversalRelationEnergy2024,onishiQuantumWeight2024,albert2016geometry,ledwith2023vortexability,yu2024non,ahn2022riemannian,kozii2021intrinsic,mckay2024charge}. 
Of particular interest is the quantum metric $G_{ij}$, which measures the polarization fluctuations in the ground state of a material at zero temperature via~\cite{marzari1997maximally,souza2000polarization} 
\begin{equation}\label{eq:quantummetricdef}
    G_{ij} = \frac{1}{V}(\langle X_iX_j\rangle - \langle X_i\rangle\langle X_j\rangle),
\end{equation}
where $X_i$ is the $i$-th component of the many-body position operator, and $V$ is the volume of the system.
For a weakly interacting system, the quantum metric  integrated over the Brillouin zone places a lower bound on how localized one can make Wannier functions for the energy bands. 
More generally, it was shown in Ref.~\cite{souza2000polarization} that the quantum metric is related via a sum rule to the zero temperature optical conductivity $\sigma^\mathrm{abs}_{ij}(\omega,T=0)$,
\begin{equation}\label{eq:conductivitysum}
    \lim_\mathbf{q\rightarrow 0}\int_0^\infty d\omega \frac{\mathrm{Re} \ \sigma^\mathrm{abs}_{ij}(\mathbf{q},\omega,T=0)}{\omega} = \frac{\pi e^2}{\hbar}G_{ij},
\end{equation}
where $\sigma^\mathrm{abs}_{ij}$ is defined in terms of the optical conductivity $\sigma_{ij}$ and the dielectric tensor $\epsilon_{ij}$ via
\begin{equation}
\sigma^\mathrm{abs}_{ij} = \frac{1}{2}\left[(\sigma\epsilon^{-1})+ (\sigma\epsilon^{-1})^\dagger\right]_{ij}.
\end{equation}
{We note that as originally written, Ref.~\cite{souza2000polarization} did not include these factors of the dielectric tensor. They implicitly computed the response to a transverse external field, for which the dielectric tensor is unity. } The combination $\sigma^\mathrm{abs}_{ij}(\omega,T=0)$ as we defined here is the (hermitian part) of the current response to an \emph{external} electric field $\mathbf{D}$ (in contrast to the optical conductivity itself, which is the response to the total electric field $\mathbf{E}$). This follows from Ohm's law
\begin{equation}
\mathbf{j}=\sigma\mathbf{E} = \sigma\epsilon^{-1}\mathbf{D}.
\end{equation}
{It is this response to an external electric field that is given directly by the Kubo formula for the conductivity, which treats the external electric field as a small perturbation. Since Ref.~\cite{souza2000polarization} derives Eq.~\eqref{eq:conductivitysum} from the Kubo formula, we have emphasized here that it is the response ot the external electric field that enters the sum rule. This subtlety is particularly important as we will consider the longitudinal conductivity below~\cite{onishiUniversalRelationEnergy2024} .}

Eq.~\eqref{eq:conductivitysum} follows from the fluctuation-dissipation theorem and relates the wavefunction geometry, encoded in $G_{ij}$, to the experimentally measurable optical conductivity. 
As such, Eq.~\eqref{eq:conductivitysum} gives a straightforward path to experimentally measuring wavefunction geometry. 
However, to our knowledge no experimental measurement of $G_{ij}$ has yet been carried out for any condensed matter system.

More recently, Ref.~\cite{onishiQuantumWeight2024} highlighted that the relationship between conductivity and density response opens a new avenue to measuring $G_{ij}$. 
In particular, conservation of charge allows one to relate the sum rule \eqref{eq:conductivitysum} to a sum rule for the zero temperature limit of the density-density response function $\chi(\mathbf{q},\omega)$.
Recall that the density-density response function is defined via the Kubo formula as
\begin{equation}
   \chi(\mathbf{q},\omega) = -\frac{i}{\hbar}\int_0^\infty dt e^{i\omega^+t} \langle \left[\rho_\mathbf{q}(t),\rho_{-\mathbf{q}(0)}\right]\rangle_0,
\end{equation}
where $\rho_\mathbf{q}$ is a Fourier component of the density operator, time evolution and averages are taken with respect to the unperturbed Hamiltonian of the system, and $\omega^+=\omega+i\epsilon$ with the limit of $\epsilon\rightarrow 0$ understood.
Using charge conservation, Ref.~\cite{onishiQuantumWeight2024} argued that asymptotically as $|\mathbf{q}|\rightarrow 0$
\begin{equation}\label{eq:chippsum}
    \lim_{T\rightarrow 0}\int_0^\infty d\omega \chi''(\mathbf{q}\rightarrow\mathbf{0},\omega) = -\frac{e^2}{2\hbar}q_iq_j G_{ij} + \mathcal{O}(|\mathbf{q}|^4),
\end{equation}
where $\chi''(\mathbf{q},\omega)$ is the imaginary part of the density-density response function. 
Through the fluctuation-dissipation theorem, $\chi''(\mathbf{q},\omega)$ can be expressed in terms of the dynamic structure factor, and so  Eq.~\eqref{eq:chippsum} shows that the quantum metric is related to the $\mathcal{O}(|\mathbf{q}|^2)$ contribution to the static structure factor at small $\mathbf{q}$ and $T=0$, also known as the quantum weight~\cite{onishiQuantumWeight2024}.
{Ref.~\cite{onishiQuantumWeight2024} takes care to distinguish the quantum weight defined in this way from the ``optical quantum weight'' which comes from integrating the negative first moment of the optical conductivity [i.e., Eq.~\eqref{eq:conductivitysum} evaluated without the factors of the inverse dielectric tensor]. As is emphasized in Ref.~\cite{resta2006polarization} and in our prior discussion, longitudinal and transverse polarization fluctuations need not coincide in systems with unscreened, long-range Coulomb interaction. The transverse components of polarization fluctuations given by the ``optical quantum weight'' give, at zero temperature, twist-angle quantum metric as shown in Ref.~\cite{souza2000polarization}. On the other hand, the longitudinal polarization fluctuations give the quantum weight of Ref.~\cite{onishiQuantumWeight2024}. Both characterize the geometry of the ground state. As we show below, the longitudinal polarization fluctuations are also related to the geometry of quantum states via the quantum Fisher information. In this work, we use $G_{ij}$ to refer to longitudinal polarization fluctuations, given by Eqs.~\eqref{eq:quantummetricdef} and Eq.~\eqref{eq:chippsum}.}

Note that while the quantum weight is expected to be finite for an insulator, it diverges for a metal; the fact that electronic states are delocalized in a metal implies that polarization fluctuations are large.
While the sum rules in Eqs.~\eqref{eq:conductivitysum} and \eqref{eq:chippsum} for the quantum metric are valid at $T=0$, we expect them to hold up to exponentially small corrections for insulating systems at nonzero temperature, provided the temperature is smaller than the optical gap. 

Inelastic X-ray scattering (IXS) is an energy- and momentum-resolved scattering technique that directly measures the electron density-electron density correlation function of a material, providing a means to determine the density response function by applying the fluctuation-dissipation theorem \cite{schuelkeElectronDynamicsInelastic2007}. 
Thus, IXS experiments on insulators, at sufficiently low temperature, can be used to measure $G_{ij}$. 
In this work, we present the first such measurement.

Further, IXS at larger values of  $\mathbf{q}$ and nonzero $T$ allows us to extract even more information about the wavefunction geometry of condensed matter systems. 
In particular, as shown in Ref.~\cite{haukeMeasuringMultipartiteEntanglement2016}, the density response $\chi''(\mathbf{q},\omega)$ determines the quantum Fisher information $f_Q(\mathbf{q},T)$ associated with density perturbations of the system,
\begin{equation}\label{eq:qfidef}
    f_Q(\boldsymbol{q},T) = -\frac{4\hbar}{\pi}\int_0^\infty d\omega\;\mathrm{tanh}\left(\frac{\hbar\omega}{2k_BT}\right)\chi''(\mathbf{q},\omega,T).
\end{equation}
The quantum Fisher information measures the rate at which the equilibrium density matrix changes under the influence of a perturbation
\begin{equation}
    \delta H(t) = \sum_{\mathbf{q}} V^\mathrm{ext}_\mathbf{q}(t) n_{-\mathbf{q}},
\end{equation}
where $V^\mathrm{ext}_\mathbf{q}(t)$ is a time-dependent external electrostatic potential and $n_{\mathbf{q}}$ is the electron density. 
If the system starts at $t=0$ in a thermal density matrix $\rho_0$, then the quantum Fisher information $f_Q(\mathbf{q},T)$ quantifies how distinguishable $\rho(t)$ is from $\rho_0$. 
In particular, the (Bures) distance $ds^2$ between $\rho_0$ and $\rho(dt)$ is given by~\cite{braunstein1994statistical}
\begin{equation}
    ds^2 = \frac{1}{4}\sum_\mathbf{q}f_Q(\mathbf{q},T)|V^\mathrm{ext}_\mathbf{q}(t=0)|^2dt^2
\end{equation}
We see then that $f_Q(\mathbf{q},T)$ quantifies the infinitesimal distance between $\rho_0$ and $\rho(dt)$ per unit intensity of the external field; in other words, $f_Q(\mathbf{q},T)$ tells us by how much the external potential can change the information content of the thermal density matrix. 
Building on this, the quantum Cramer-Rao bound implies that the distance $ds^2$---and hence $f_Q(\mathbf{q},T)$---is the limit on the precision with which we can ever distinguish $\rho_0$ from $\rho(dt)$ through any set of measurements.

The quantum Fisher information \eqref{eq:qfidef} extends the sum rule \eqref{eq:chippsum} both to finite wavevector and nonzero temperature. 
In particular, we can examine the low-temperature limit of $f_Q(\mathbf{q},T)$. 
For an insulator at low temperatures, the susceptibility $\chi''(\mathbf{q},\omega)$ vanishes up to exponentially small correction for $\hbar\omega$ smaller than the optical gap. 
Furthermore, as $T\rightarrow 0$ the $\mathrm{tanh}$ in Eq.~\eqref{eq:qfidef} becomes exponentially close to $1$. 
We thus have for an insulator that
\begin{equation}
    f_Q(\mathbf{q},T\rightarrow 0) \approx -\frac{4\hbar}{\pi}\int_0^\infty d\omega \chi''(\mathbf{q},\omega).
    \label{eq: TZeroLimit}
\end{equation}
Combining this with Eq.~\eqref{eq:chippsum}, we find that at low temperatures and small $\mathbf{q}$, the quantum Fisher information for an insulator is given by the quantum weight. 
Note that for a metal, this correspondence breaks down; the linearity of $\mathrm{tanh}$ for small arguments ensures that the QFI for a metal is finite even when the sum rule \eqref{eq:chippsum} diverges. 
Lastly, we note that at high temperatures, the QFI exhibits universal behavior dictated by the $f$-sum rule. 
In particular, as $T\rightarrow\infty$ we can Taylor expand the $\mathrm{tanh}$ in Eq.~\eqref{eq:qfidef} to find
\begin{align}
    f_Q(\mathbf{q},T\rightarrow\infty)&\sim-\frac{2\hbar}{\pi k_BT}\int_0^\infty d\omega\; \omega\chi''(\mathbf{q},\omega) \\
    &\sim \frac{1}{k_B T}\left(\frac{\hbar n e^2}{m}\right),
\end{align}
where the integral is fixed by the $f$-sum rule. 
Thus, as temperature increases, the amount of information contained in the perturbed density matrix $\rho(dt)$ decays to zero as $1/T$ with a coefficient determined by the density of particles. We thus see that the quantum Fisher information generalizes quantum geometry to finite temperature and to systems that are not insulating.

There is thus a tremendous need for experimental techniques capable of measuring the quantum metric and quantum Fisher information and quantifying the above relationships. 
Recently, an inelastic neutron scattering study of the 1D spin chain KCuF$_3$ showed that the dynamic spin susceptibility can be used to determine the quantum Fisher information, where it may be interpreted as a measure of the number of entangled spins, the results being in excellent agreement with expectations from Bethe ansatz and DMRG techniques \cite{scheieWitnessingEntanglementQuantum2021}. 
This approach was also used to analyze spin excitations in the 3D iridate dimer material Ba$_3$CeIr$_2$O$_9$ using resonant inelastic X-ray scattering (RIXS) \cite{Ren2024}, and to study the spin entanglement of strange metals~\cite{fang2024amplified,mazza2024quantum}.
There is still, however, a need to apply this approach using the charge response, which would allow analysis of continuum systems, such as metals, and testing bounds of the sort predicted in Ref. \cite{onishiQuantumWeight2024} through a direct measurement of the quantum weight. 
Here, we present such a study using the charge response measured with inelastic X-ray scattering (IXS). 

The key result of Ref.~\cite{onishiQuantumWeight2024} is a fundamental bound of the quantum weight as a function of the system's band gap and dielectric constant as well as computed bounds for common insulating systems. 
Among the insulators considered in Ref.\cite{onishiQuantumWeight2024} was lithium fluoride (LiF). 
In previous work, we performed IXS measurements on LiF which showed a well defined band gap and Frenkel exciton \cite{abbamonteDynamicalReconstructionExciton2008}. 
This work provides an opportunity to experimentally test the proposed bounds on the quantum weight. 

In what follows, we will first review the experimental details from Ref.~\cite{abbamonteDynamicalReconstructionExciton2008}. 
We will then show how we can use the IXS data obtained in that experiment to compute both the quantum Fisher information and quantum weight of LiF. 
We will then compare our result to the bounds presented in Ref.~\cite{onishiQuantumWeight2024}, and conclude with an outlook towards future applications of IXS to quantum geometry and quantum entanglement.

\section{Experiment}

The IXS data from LiF analyzed in this study were previously reported in Ref. \cite{abbamonteDynamicalReconstructionExciton2008}. 
Briefly, the differential scattering cross section for IXS is given by \cite{schuelkeElectronDynamicsInelastic2007}

\begin{equation}
    \frac{\partial^2 \sigma}{\partial \Omega \partial \omega} = r_0^2 \frac{\omega_f}{\omega_i} \left | \epsilon_f^* \cdot \epsilon_i \right |^2 S(\mathbf{q},\omega)
\end{equation}

\noindent where $r_0$ is the classical electron radius, $\epsilon_i$ and $\epsilon_f$ are the incident and scattered polarizations, respectively, and $\omega_i / \omega_f \approx 1$ is the ratio of incoming and outgoing photon energies. ${\bf q}$ and $\omega$ represent, respectively, the momentum and energy transferred to the material during the scattering event. 
The van Hove function, $S({\bf q},\omega)$, is the density-density correlation function of the material, also known as the dynamic structure factor. 
It is related to the dynamic charge susceptibility by the quantum mechanical version of the fluctuation-dissipation theorem,

\begin{equation}\label{eq: flucDisThrm}
    \chi''({\bf q},\omega) = -\frac{1}{\pi} \frac{1}{1-e^{-\hbar \omega/k_BT}} S(\mathbf{q},\omega).
\end{equation}

\noindent Hence, up to an overall multiplicative constant, IXS directly measures $\chi''(\mathbf{q},\omega)$, which is of interest for computing the quantum Fisher information and quantum metric via Eqs.~\eqref{eq:qfidef} and \eqref{eq:chippsum}, respectively. 
The multiplicative constant can be determined using a sum rule, as illustrated below. 

All IXS data were taken at room temperature, $k_B T \sim 25$ meV, which is small compared to the (optical) band gap of LiF, $E_g \sim 15$ eV \cite{abbamonteDynamicalReconstructionExciton2008}. 
Hence, the Bose factor in Eq.~\eqref{eq: flucDisThrm} can be approximated as $[1-e^{-\hbar \omega/k_BT}]^{-1} \approx 1$ for all relevant frequencies. 
The momentum transfer in the experiment, $\boldsymbol{q}$, was parallel to the $(h,k,l) = (1,0,0)$ direction with magnitude $q=|\mathbf{q}|$ ranging from $0.318 \  \text{\AA}^{-1} \le q \le 3.7 \ \text{\AA}^{-1}$. 
For each $q$, the energy loss was scanned from $-10 \ eV$ to $100 \ eV$.

\begin{figure}
	\center
	\includegraphics[width=1\linewidth]{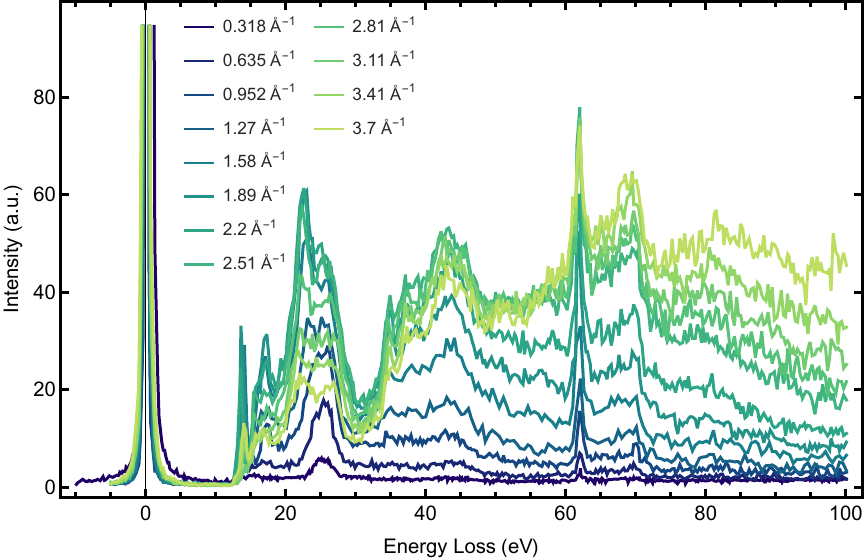}
	\caption{IXS spectra, which are proportional to $\chi''(\mathbf{q},\omega)$, from LiF measured at the Advanced Photon Source. 
    Each color corresponds to a different momentum transfer in the $(100)$ direction. 
    The vertical scale is in photons/sec. 
    These spectra can be rescaled to the units of $\chi''$ by using the f-sum rule (see text). }
	\label{fig: rawData}
\end{figure}

Raw IXS spectra, in units of photons/sec, are shown in Fig. \ref{fig: rawData}. 
The main spectral features are a strong elastic peak at $\omega = 0$, a pronounced exciton at $\omega=14.5$ eV, interband transitions for $\omega > 16$ eV, a plasmon-like feature at $\omega \sim 25$ eV, as well as some shallow core levels and the Li K edge, which appears at $\omega \sim 60$ eV. 
The broad background that emerges at larger values of $q$ represents the emergence of a Compton profile \cite{abbamonteDynamicalReconstructionExciton2008}. 
These data should be suitable for determining the quantum weight and testing the bound predicted in Ref. \cite{onishiQuantumWeight2024}. 
First, we will need to use the $f$-sum rule to calibrate the  vertical scale in Fig.~\ref{fig: rawData} to determine the absolute magnitude and units for $\chi''$.

\section{Computing the Quantum Weight}

Our goal is to use this IXS data to compute the quantum Fisher information $f_Q(\mathbf{q},T)$ and ultimately the quantum weight. Since the wavevector used in the IXS experiment $\mathbf{q}=q(1,0,0)$ is parallel to the $(h,k,l)=(1,0,0)$ direction, we can us the IXS data to compute~\cite{onishiQuantumWeight2024}
\begin{equation}\label{eq:KfromGij}
    K=G_{11}
\end{equation}
given by the component of the quantum metric $G_{ij}$ parallel to the wavevector, per Eq.~\eqref{eq:chippsum}~\footnote{Other components of the metric $G_{ij}$ could be obtained by considering wavevectors oriented along different directions. For instance $\mathbf{q}=q(1,1,0)$ would yield, according to Eq.~\eqref{eq:chippsum}, the linear combination $G_{11}+G_{22}+2G_{12}$}.
%\dbnote{we bring this in kind of out of the blue here. Maybe we should mention in the introduction that an alternative relation for coulomb systems can be written} 
In this case, the quantum weight is given by a sum over $\chi''$ with Eq.~\eqref{eq:chippsum} as well as a weighted sum over the (longitudinal) optical conductivity and the (longitudinal) dielectric function~\cite{onishiQuantumWeight2024} via Eq.~\eqref{eq:conductivitysum} as:
\begin{equation}
\lim_{q\rightarrow 0}\int_{0}^{\infty}d\omega \frac{1}{\omega} \ \text{Re} \left[\sigma(\mathbf{q},\omega)\epsilon^{-1}(\mathbf{q},\omega)\right]_{11} = \frac{e^2}{2 \hbar}K   
\label{KixsDef}
\end{equation}
where we have used the fact that the dynamic charge response is related to the longitudinal optical response through~\cite{nozieres1999}, 

\begin{equation}
    \frac{\sigma(\mathbf{q},\omega)}{\epsilon(\boldsymbol{q},\omega)}\equiv [{\sigma(\boldsymbol{q},\omega)\epsilon^{-1}(\boldsymbol{q},\omega)}]_{11} = -i \frac{ e^{2}\omega}{q^2}\chi(\boldsymbol{q},\omega),
	\label{eq: sigmaDefinition}
\end{equation} 
which is a statement of charge conservation (i.e., the continuity equation) in frequency- and momentum-space. 
Making use of Eq.~\eqref{eq: sigmaDefinition}, we can write the quantum Fisher information for an insulator at low temperatures $k_BT \ll E_g$ as
\begin{equation}\label{eq:fqfromsigmaq}
    f_Q(\mathbf{q},T\rightarrow 0) = \frac{4\hbar q^2}{\pi e^2}\int_0^\infty d\omega \frac{1}{\omega}~\mathrm{Re}~\frac{\sigma(\mathbf{q},\omega)}{\epsilon(\boldsymbol{q},\omega)}.
\end{equation}
Using Eqs.~\eqref{eq: sigmaDefinition} and \eqref{eq: flucDisThrm}, we can apply Eq.~\eqref{eq:fqfromsigmaq} to compute $f_Q(\mathbf{q},T\rightarrow 0)$ directly from IXS data. 

\begin{figure}
	\centering
	\includegraphics[width=1\linewidth]{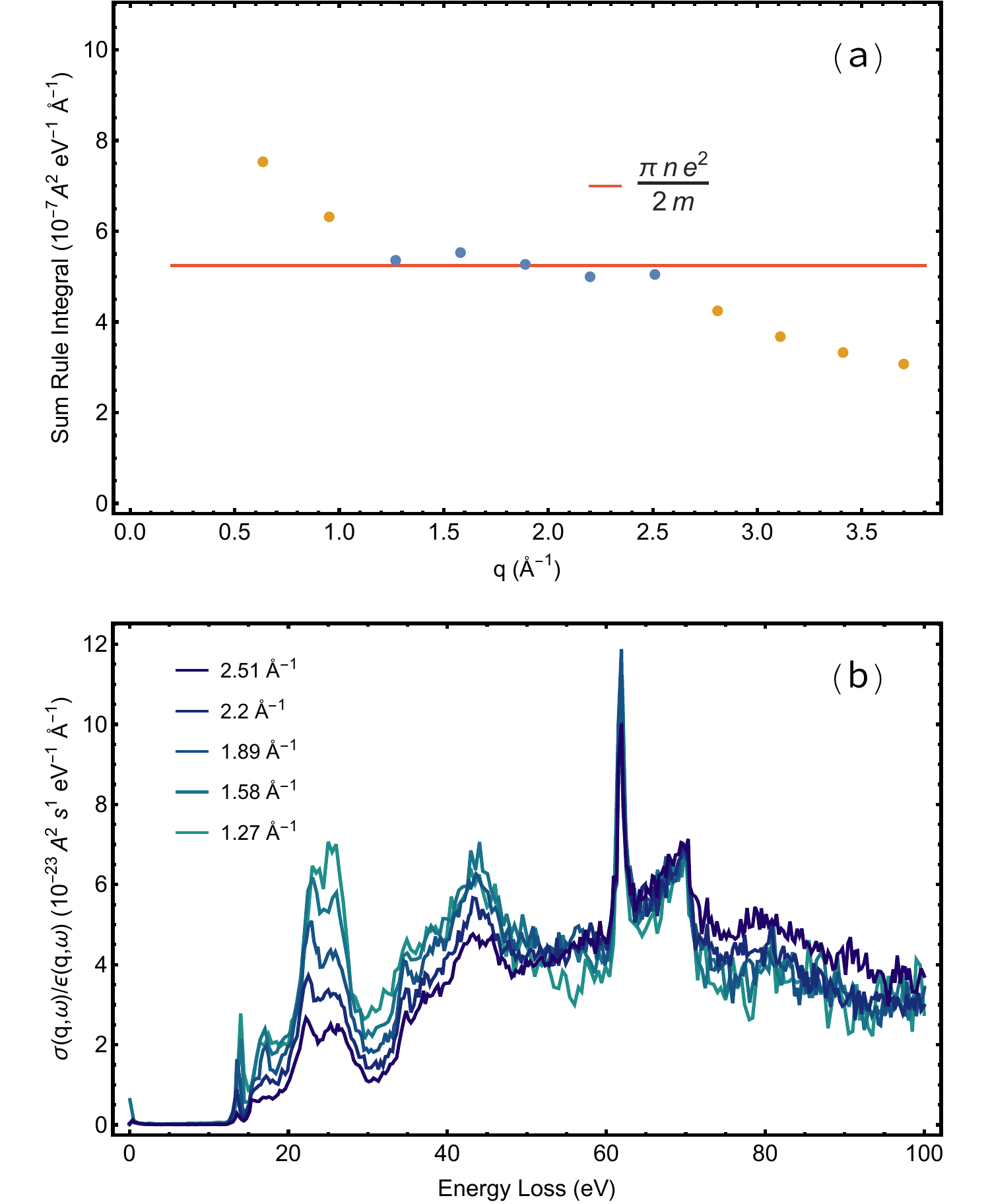}
	\caption{a) Sum rule integral value from Eq.~\eqref{eq:sigmaSumRule}. 
    The spectra were scaled with a $q$-independent constant. 
    Values of momentum transfer that follow the sum rule are plotted in blue whereas those that do not are plotted in orange. b) Real part of $\sigma(\mathbf{q},\omega)/\epsilon(\boldsymbol{q},\omega)$  computed from Eq.~\eqref{eq: sigmaDefinition} for various momenta.}
	\label{fig:sigmaPlots}

\end{figure}

To make use of Eqs.~\eqref{eq: sigmaDefinition} and \eqref{eq:fqfromsigmaq}, however, we need to fix the overall scale of the IXS data. 
This can be accomplished by applying the $f$-sum rule \cite{kuboStatisticalMechanicalTheoryIrreversible1957,kadanoff1963hydrodynamic},
\begin{equation}
	\label{eq:sigmaSumRule}
	\begin{split}
		\int_{0}^{\infty}d\omega \ \mathrm{Re}  \frac{\sigma(\boldsymbol{q},\omega)}{\epsilon(\boldsymbol{q},\omega)} 
		&= \frac{-e^2}{q^2} \int_{0}^{\infty} d\omega \ \omega \ \mathrm{Im} \chi(\boldsymbol{q},\omega) \\
		&= \frac{\pi n e^2}{2m}.
	\end{split}
\end{equation}

\noindent where $n$ is the equilibrium electron density and $m$ is the electron mass. 
For LiF, $n$ is reported to be $0.739 \ \text{\AA}^{-3}$ \cite{onishiUniversalRelationEnergy2024}. 
In Fig.~\ref{fig:sigmaPlots} we show the conductivity computed from the raw IXS data and scaled by a single momentum-independent constant to enforce the sum rule \eqref{eq:sigmaSumRule}. 
The constant was found by first numerically integrating the spectra from $\hbar \omega  = 0~\text{eV}$ to $\hbar \omega = 100~\text{eV}$. Then, the values of the $f$-sum integral for momenta in the range $1.27 \text{\AA}^{-1} \le q \le 2.51 \text{\AA}^{-1}$ , where the value of the sum rule is independent of momentum, were averaged. Finally, these spectra were scaled with a constant to match the rightmost side of Eq.~\ref{eq:sigmaSumRule}. This was done under the assumption that the effective electron density from $0$ to $100$ eV is $n = 0.739 \ \text{\AA}^{-3}$. 
Experimentally, the IXS spectra at very small $q$ contain extra background signal due to interference from the forward-scattered beam. 
At large $q$, spectral weight ``leaks" out of the scan region due to the emergence of Compton scattering for energy losses greater than 100 eV. 
Both these effects can be seen in the deviation from the $f$-sum rule in Fig.~\ref{fig:sigmaPlots}(a). 
However, for momenta in the range $1.27 \text{\AA}^{-1} \le q \le 2.51 \text{\AA}^{-1}$, these two effects are minimal, and a single, $q$-independent constant scales the spectra to obey Eq.~\eqref{eq:sigmaSumRule}, allowing a reliable measure of the quantum weight and quantum Fisher information for momenta in this range. The dimensionful optical conductivity spectra determined by this means are shown in Fig.~\ref{fig:sigmaPlots}(b).  
%\dbnote{Here I try to clarify how we found the scaling constant since one of the reviewers commented on it. I don't like how it reads but left it here for your opinions}

\begin{figure}
	\centering
	\includegraphics[width=1\linewidth]{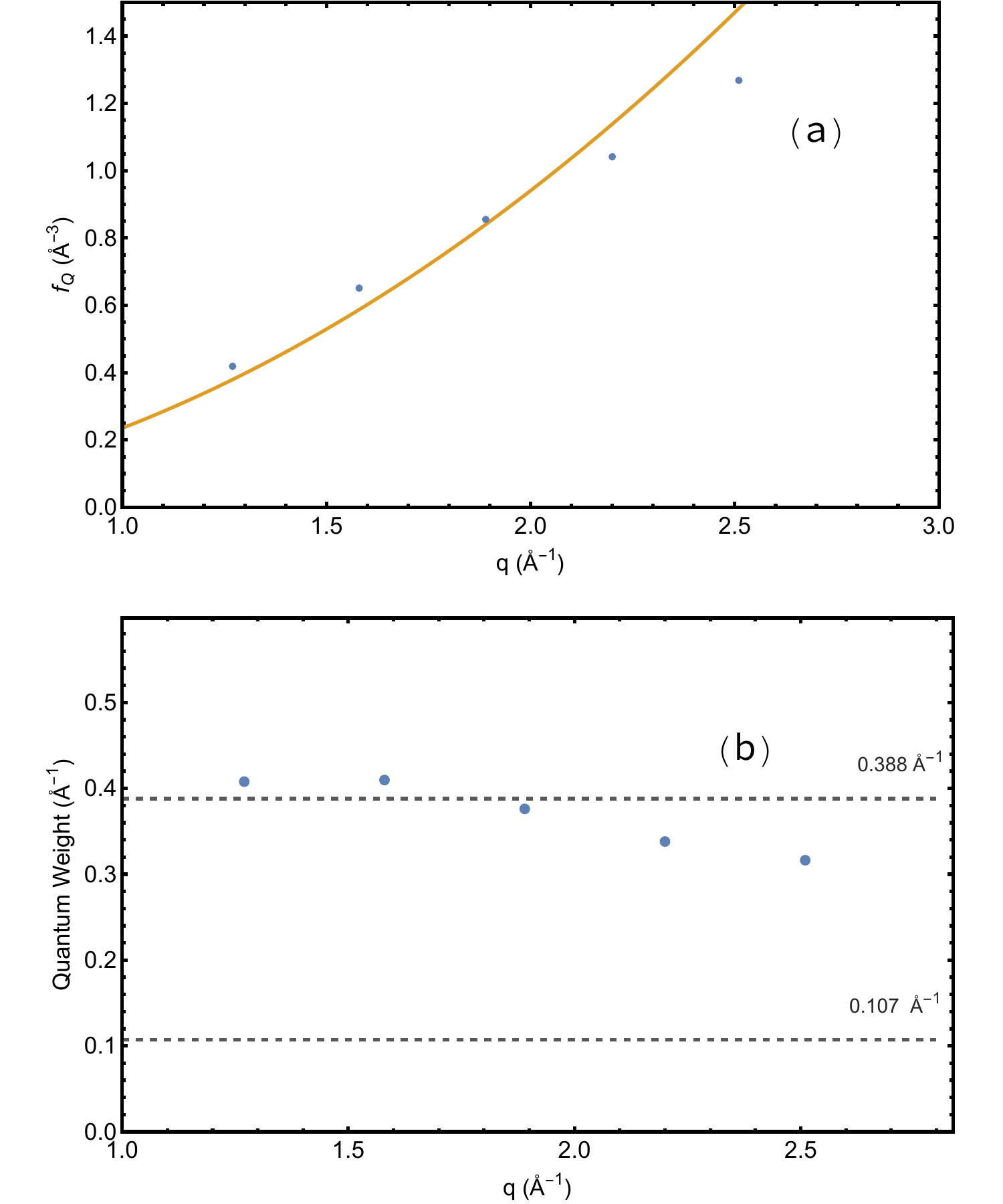}
\caption{Quantum Fisher information and quantum weight for LiF. (a) shows the quantum Fisher information computed from Eq.~\eqref{eq: TZeroLimit}. (b) shows the sum rule \eqref{eq:chippsum} divided by $q^2$ to extrapolate the quantum weight to $q\rightarrow 0$. 
The horizontal dashed lines represent the theoretical upper and lower bounds from Eq.~\eqref{eq: quantumWeightBound} using the material parameters of LiF. 
The extrapolated value for the quantum weight was found using Eqs.~\eqref{eq: TZeroLimit}, \eqref{qfiAndK} and averaging the result over q. The result is $K\approx 0.37 \pm 0.04 \AA^{-1}$.
%$K\approx 0.369589 \ \text{\AA}^{-1}$. 
The yellow curve in (a) is the approximate quadratic expression for $f_Q$ from Eq.~\eqref{qfiAndK} using the extrapolated value for $K$. 
For computing both $K$ and $f_Q$ we focused on the five values of $q$ for which the $f$-sum rule was satisfied in the experiment (see Fig.~\ref{fig:sigmaPlots}), and made use of the fact that $\chi''(q,\omega)\approx 0$ for $\hbar\omega < E_g$ in LiF at low temperature to simplify the frequency integrals.}
	\label{fig: quantumWeightPlot}
\end{figure}

In Fig.~\ref{fig: quantumWeightPlot}(a) we show the quantum Fisher information $f_Q(\mathbf{q},T\rightarrow 0)$ for LiF computed from IXS data at the five wavevectors for which the $f$-sum rule is satisfied. 
We see that the $q$-dependence of the Fisher information is approximately quadratic, suggesting that the wavevector dependence of the conductivity in Eq.~\eqref{eq:fqfromsigmaq} can be ignored. 
To explore this further, we note that we can take the small $\mathbf{q}$ limit of our approximation Eq.~\eqref{eq: TZeroLimit} for the quantum Fisher information for insulators at low temperatures, combined with the small $\mathbf{q}$ approximation Eq.~\eqref{eq:chippsum} for $G_{ij}$ to deduce that for insulators
\begin{equation}
    f_Q(\mathbf{q}\rightarrow 0,T\rightarrow 0) = \frac{2}{\pi}q^{2}K.
    \label{qfiAndK}
\end{equation}
With this as motivation, we plot $\pi f_Q(q,T\rightarrow 0) / 2q^2$ obtained via IXS as a function of $q$ in Fig.~\ref{fig: quantumWeightPlot}, for the five $q$ points at which the $f$-sum rule is satisfied. 
We see that the data are approximately $q$-independent, allowing us to extrapolate to $q\rightarrow 0$, average the resulting values, and identify this result with the quantum weight $K \approx 0.37 \pm .04 \ \text{\AA}^{-1}$. 
As a sanity check, we can also compute the quantum weight by first determining the optical conductivity via Eq.~\eqref{eq: sigmaDefinition}, and then integrating using Eq.~\eqref{eq:fqfromsigmaq} to get $f_Q/q^2$. 
Extrapolating to $q=0$ yields $K\approx 0.36 \pm .04 \ \text{\AA}^{-1}$, in statistically equivalent to the direct approach.

We now wish to use this number to test the central claim of Ref. \cite{onishiQuantumWeight2024}, which is a universal bound on the quantum weight,
\begin{equation}
	\pi(1 - \epsilon^{-1})E_g \leq \frac{e^2 K}{\epsilon_0} \leq \pi\sqrt{1 - \epsilon^{-1}} \hbar \omega_p
	\label{eq: quantumWeightBound}
\end{equation}
where $E_g = \hbar \omega_{g}$ is the band gap of the system, $\epsilon$ is the electronic contribution to the static dielectric constant, and $\omega_p \equiv \sqrt{\frac{n e^2}{m \epsilon_0}}$ is the ``bare plasma frequency".
These bounds are shown in Fig. \ref{fig: quantumWeightPlot}(b), where they are depicted as horizontal dashed lines, where we use the values of $\epsilon$,$n$ are reported in Ref.~\cite{onishiQuantumWeight2024}.  
The quantum weight lies within the bounds given by Eq.~\eqref{eq: quantumWeightBound}, falling near the upper bound.

The definition Eq.~\eqref{eq:KfromGij} of the quantum weight shows that it is given by the magnitude of ground state polarization fluctuations via Eq.~\eqref{eq:quantummetricdef} defining the quantum metric. This result suggests quantum mechanical fluctuations are important even in an ionic insulator. 

\section{Discussion}

In this work we have shown how inelastic X-ray scattering may be used to reveal the quantum information and quantum geometry encoded in the ground state of materials, in this case the large-gap insulator, LiF. 
We found that, following the proposal of Ref.~\cite{onishiQuantumWeight2024}, the experimentally measured susceptibility $\chi''(\mathbf{q},\omega)$ can be integrated to obtain the quantum weight $K$, which is intimately connected to the quantum metric and hence the wavefunction localizability in LiF. 
In particular, $K$ is proportional to the square of the localization length for electrons in the ground state of a solid. 
We determined experimentally that the quantum weight in LiF lies close to the upper bound determined from the energy gap and the static charge susceptibility. 
This suggests that, while LiF is a very strongly localized ionic insulator, electrons are nearly as delocalized as they are theoretically allowed to be. More precise measurements of the quantum weight using more modern spectroscopic techniques could shed further light on this issue.

Additionally, we evaluated the quantum Fisher information $f_Q(\mathbf{q},T)$ from the measured susceptibility. 
We argued that for large gap insulators like LiF, the room-temperature quantum Fisher information is approximately equal to the integrated susceptibility, and hence at small $\mathbf{q}$ we showed that the quantum Fisher information scales approximately as $|\mathbf{q}|^2$ with a coefficient given by the quantum weight via Eq.~\eqref{qfiAndK}. Furthermore, we have shown how the quantum Fisher information provides a natural path to generalize the quantum weight (and quantum geometry more generally) to systems at nonzero temperature as well as to systems without an optical gap.

Our work serves as a proof of principle that inelastic X-ray scattering is a powerful tool for probing quantum geometry and entanglement in materials. 
Localization and geometry in insulators are normally thought of as electrostatic properties, most easily probed via the AC conductivity. 
Here, we have shown, following the suggestion of Refs.~\cite{onishifundamentalbound,onishiQuantumWeight2024}, not only that IXS can yield an accurate measurement of the quantum weight, but also that it allows for the measurement of the $\mathbf{q}$ and $\omega$ dependent conductivity via Eq.~\eqref{eq: sigmaDefinition}. 
These results also highlight the crucial role that sum rules play in the interpretation of spectroscopic data: Not only is the quantum weight defined directly via a sum rule, but the $f$-sum rule is essential in calibrating the scale of any spectroscopic measurement. 

We have also revealed a deep connection between localization (as encoded in the quantum weight $K$) and quantum information, measured via the quantum Fisher information. 
The quantum Fisher information is directly proportional to the degree of entanglement present in the ground state of the system, as it tells us by how much a density perturbation can change the state. 
We have seen, via Eq.~\eqref{qfiAndK} and our experimental results, that the more localized a system is, the smaller $K$, and hence the smaller the quantum Fisher information. 
This matches with our intuition that strongly-bound ionic systems like LiF should have trivial ground states and hence a low degree of entanglement. 
Our work takes this beyond the realm of intuition and gives a concrete experimental prescription for measuring the information content of a real material. 
That this is possible is perhaps surprising, given that any condensed matter system has $\mathcal{O}(10^{23})$ electrons that all participate in absorbing energy in any spectroscopic experiment. 
Nevertheless, the rigid constraints placed by the $f$-sum rule and the sum rule \eqref{eq:conductivitysum} ensure that the quantum Fisher information is measurable.

Looking forward, our work motivates several areas for further research. 
First, we chose to focus here on LiF as it is the simplest system that illustrates how $f_Q$ and $K$ can be computed. Because of its simplicity, one area of future work could be to attempt to compute $\chi(\mathbf{q},\omega)$ and $f_Q$ for LiF directly using ab initio or other approximate techniques. This is a difficult problem even for such a simple material, where one must go beyond RPA to fully capture correlation and excitonic effects~\cite{caliebe2000dynamic,reed2010effective,husain2023pines,kengle2023non}.

Next, it would be interesting to apply our analysis technique to nontrivial systems such as metals (where $K$ diverges although we expect $f_Q$ to remain finite), topological insulators, and doped Mott insulators. 
For such systems, a direct measure of both the localization length via $K$ and the quantum information via $f_Q$ could shed light on their nontrivial dynamics. 
Furthermore, we note that while we used IXS as our measurement tool, an alternative approach is to use momentum resolved inelastic electron scattering (M-EELS). This also measures the charge response, but with much higher resolution, close to $2 meV$, which enables one to study low-energy excitations in more exotic materials, such as superconductors, topological materials, and other strongly interacting systems~\cite{abbamonteCollectiveChargeExcitations2024,vigMeasurementDynamicCharge2017}. M-EELS is also a surface technique, allowing it to be applied to 2D systems for which IXS is inappropriate, such as graphene and twisted van der Waals materials. This would allow for an experimental probe of the interplay between quantum geometry and topology in these systems~\cite{torma2022superconductivity}.

Lastly, we have focused here on density-density correlation functions due to their relationship to the quantum weight and Fisher information. 
However, recent work~\cite{tam2022topological,tam2024topological} has shown that for Fermi liquids, multipartite entanglement is encoded in three- and higher-point density correlation functions. 
Extending X-ray spectroscopic techniques and theoretical analysis~\cite{bradlyn2024spectral,watanabe2020generalized,matsyshyn2019nonlinear} to measure these higher-order correlation functions is a promising avenue to learn even more about the quantum information of condensed matter systems.

\section{Acknowledgments}

This work was primarily supported by the Center for Quantum Sensing and Quantum Materials, an Energy Frontier Research Center funded by the U.S. 
Department of Energy (DOE), Office of Science, Basic Energy Sciences (BES), under award DE-SC0021238. 
P.A. gratefully acknowledges additional support from the EPiQS program of the Gordon and Betty Moore Foundation, grant GBMF9452. 
The theoretical work of B.B. on wavefunction geometry was additionally supported by the Alfred P. 
Sloan foundation, and the National Science Foundation under grant DMR-1945058. 
IXS measurements at the Advanced Photon Source were supported by DOE grant DE-AC02-06CH11357. 
D.B. was supported in part by the A.C. 
Anderson Undergraduate Research Scholar Award, Department of Physics, University of Illinois Urbana-Champaign.

\bibliography{References}

\end{document}